\documentclass[graybox, nosecnum]{svmult}

\usepackage{amssymb}
\usepackage{mathptmx}       
\usepackage{helvet}         
\usepackage{courier}        
\usepackage{type1cm}        
\usepackage{makeidx}         
\usepackage{graphicx}        
\usepackage{multicol}        
\usepackage[bottom]{footmisc}
\usepackage{hyperref}        
\usepackage{soul}            
\hypersetup{colorlinks=true,urlcolor=blue}
\usepackage[square,numbers]{natbib}
\makeindex             

\newcommand{\comment}[1]{{\bf \color{red} #1}}
\newcommand{\commentout}[1]{}

\begin{document}
\title*{Detecting gamma rays with high resolution and moderate field of view: the air Cherenkov technique}
\titlerunning{IACTs} 
\author{Juan Cortina and Carlos Delgado 
}

\institute{Juan Cortina \at CIEMAT, Avda. Complutense 40, Madrid, Spain, \email{juan.cortina@ciemat.es}
\and Carlos Delgado \at CIEMAT, Avda. Complutense 40, Madrid, Spain, \email{carlos.delgado@ciemat.es}
}
%
%
\maketitle
\abstract{
The Imaging Atmospheric Cherenkov technique allows to detect very high energy gamma rays from few tens of GeV to hundreds of TeV using ground-based instrumentation. At these energies a gamma ray generates a shower of secondary particles when it enters the Earth's atmosphere. These particles emit Cherenkov light in the visible and near UV ranges. The Cherenkov light produced by the shower reaches the ground as a short pulse of a few nanosecond duration over a large circle of around 100 m radius (a light pool). This pulse of light can be imaged with telescopes provided with fast photodetectors and electronics. Combining the images of several telescopes distributed over this light pool allows to estimate the gamma-ray energy and incident direction, and to reject gamma rays from the strong background of charged cosmic rays. The collection area of an array of a few telescopes is of the order of the area of the light pool, i.e. $>$10$^5$m$^2$. Such an array reaches a sensitivity of a few millicrabs at 100~GeV energies in 50~hours of observations, an angular resolution of $\sim$5~arcmin and a spectral resolution of $\sim$10\%. This chapter describes the technical implementation of Imaging Atmospheric Cherenkov telescopes and describes how the data are analyzed to reconstruct the physical parameters of the primary gamma rays. 
}
\section{Keywords} 
IACTs, Cherenkov Telescopes, very high energy gamma rays, cosmic rays, instrumentation

\commentout{Mazin:I think it would be good to add:
\begin{itemize}
\item Sensitivity calculation and Sensitivity plot
\item Example of a detection plot
\item Example of an energy spectrum
\item Example of a sky map for a point like source and an extended gamma-ray source
\item Example of a light curve
\end{itemize}

I think it might be good to discuss the future possibilities and limitations of the IACT technique. Maybe a short paragraph at the end?
}

\section{Introduction}
\commentout{Introduction to the chapter; length depends on the topic describing importance of subject and content.}


The imaging atmospheric Cherenkov technique was pioneered by the Whipple Collaboration in the USA\cite{handbook_history}.  After more than 20 years of development, Whipple discovered the Crab nebula, the first VHE gamma-ray source, in 1989\cite{Whipple}. The Crab nebula is one of the most powerful sources of very high energy gamma rays, and is often used as a "standard candle". Modern instruments, which use multiple telescopes to follow the cascades from different perspectives and employ fine-grained photon detectors to enhance the images, can detect sources with a flux below 1\% of the Crab Nebula flux. Finely pixelated images were first employed on the French CAT telescope\cite{CAT}, and the use of "stereoscopic" telescope systems to provide images of the cascade from different viewpoints was pioneered on the European HEGRA IACT system\cite{HEGRA}. \commentout{For summaries of the achievements of recent years and the scientific case for a next-generation very-high-energy gamma-ray observatory, see \cite{Reports}.}

Irrespective of the technical implementation details, as far as its performance is concerned, an Imaging Atmospheric Cherenkov Telescope (IACT) is primarily characterised by its light collection capability, i.e. the product of mirror area, photon collection efficiently and photon detection efficiency, by its field of view and by its pixel size, which limits the size of image features which can be resolved. The larger the light collection efficiency, the lower the gamma-ray energy that can be successfully detected. The  optical  system of the telescope should obviously be able at achieve a point spread function matched  to  the  pixel  size. The  electronics  for  signal  capture  and  triggering should  provide  a  bandwidth  matched  to  the  length  of  Cherenkov  pulses  of a  few  nanoseconds.  The  performance is also  dependent  on  the triggering strategy; Cherenkov emission from air showers has to be separated in  real  time  from  the  high  flux  of  night  sky  background  photons,  based  on individual images and global information in case the showers are observed from several viewpoints. In addition, the huge data stream from IACTs does not allow to deal with untriggered recording easily.

The collection area of an array of 2-4 telescopes is of the order of the area of the light pool, i.e. $>$10$^5$m$^2$. This collection area may grow by simply adding more telescopes to the array. An array of a 2-4 telescopes of the 10~m diameter mirror class reaches a sensitivity of a few millicrabs at 100~GeV energies in 50~hours of observations, an angular resolution of $\sim$5~arcmin and a spectral resolution of $\sim$10\%. Larger mirrors (20-30 m) bring the energy threshold of the array to few tens of GeV while increasing the number of 10~m diameter telescopes to 10-20 can improve the sensitivity well under 1 millicrab. However telescope optics prevents the field-of-view (FOV) of IACTs to exceed 8-10$^{\circ}$ diameter and the Cherenkov light of atmospheric showers can only be detected during the astronomical night, typically without Moon, and with good weather conditions, so the duty cycle of IACTs rarely exceeds 15\%.

These collection areas are orders of magnitude larger than the collection areas of satellite-based detectors\cite{handbook_space}. This makes IACTs the instruments of choice in the energy range between a few tens of GeV and tens of TeV. During the last two decades IACTs have opened a new astronomical window: the VHE $\gamma$-ray range offers a new tool to study sources of non-thermal radiation such as supernova remnants, star forming regions or the surroundings of compact objects (jets and winds around black holes and pulsars). VHE $\gamma$-ray astronomy also allows to study the extragalactic background light and intergalactic magnetic fields or to search for dark matter and effects of quantum gravity\cite{handbook_science}.

At larger energies, from hundreds of TeV to a few PeV, one requires even larger collection areas, larger FOVs and/or a duty cycle approaching 100\%, as offered by detector arrays sampling the particles in the atmospheric shower. However IACTs still offer unbeatable angular and spectral resolutions.

\section{Air shower properties and imaging}\label{sec:air shower}

%
%
%
%
%
%

When a very high energy gamma-ray interacts with the Earth's upper atmosphere, it converts into an electron-positron pair. Subsequent Bremsstrahlung and pair production interactions generate an electromagnetic shower in the atmosphere, in which the total number of electrons, positrons and photons approximately doubles every $log(2)$ times the radiation length for Bremsstrahlung in the atmosphere (roughly 37.2 $g\ cm^{-2} $)\cite{Heitler}. The splitting of the energy of the primary gamma-ray stops when the losses due to ionization of the secondary electrons and positrons dominate over the other processes. This happens when the average lepton energy is about $E_0=84\ MeV$ and, as a result, the maximum number of leptons in the shower is $E/E_0$, where $E$ is the energy of the primary.

Given their energy, the electrons and positrons move faster than the speed of light in air, thus emitting Cherenkov radiation\cite{Cherenkov light}. The maximum intensity of this emission occurs when the number of particles in the cascade is largest, at an altitude of $\sim$10 km for primary gamma-ray energies of 100 GeV to 1 TeV near the zenith. During their propagation these particles undergo multiple Coulomb scattering, distributing then in the direction perpendicular to the shower propagation, increasing the spread of their individual direction of propagation. This together with the Cherenkov angle, which amounts to about $1.4^\circ$, result of a pool of photons nearly uniformly distributed within a circle of about 130~m of radius around the extrapolation of the primary trajectory to ground,  with a density of about 100 photons m$^{-2}$s$^{-1}$ for a primary gamma ray of about 1 TeV.\commentout{, as illustrated in Fig. \ref{fig:pool},} These photons arrive to the ground in a single pulse of few nanoseconds of duration.
\commentout{
\begin{figure}
    \centering
    \includegraphics{}
    \caption{Photon density from electromagnetic showers obtained from Flux simulations, for 500 GeV and 100 GeV primary energy.}
    \label{fig:pool}
\end{figure}
}

An IACT detects these photons and determines and kinematics of the primary. The detection technique is based on a simple concept: photons are collected in a large mirror which focuses them in a fast camera with photodetectors coupled to digital samplers\footnote{Actually the first detection of Cherenkov from an air shower was done with a free running analog oscilloscope and a single photomultiplier.}. However, as usual, the devil is in the details: even if detection is relatively simple, rejection of cosmic ray showers, night sky photons and determination of the gamma-ray kinematics are challenging problems. 

Cosmic rays (mainly protons and He nuclei) collide with atmospheric nuclei to generate secondary hadronic or leptonic particles. After further interactions they develop several electromagnetic sub-showers and muons. Compared to a gamma-ray shower and due to the larger transverse momentum of hadronic interactions, cosmic-ray shower particles spread away from the incident direction. The corresponding shower image at an IACT is broader and more irregular. This fact is key to cosmic-ray rejection techniques.

\begin{figure}
    \centering
    \includegraphics[width=\textwidth]{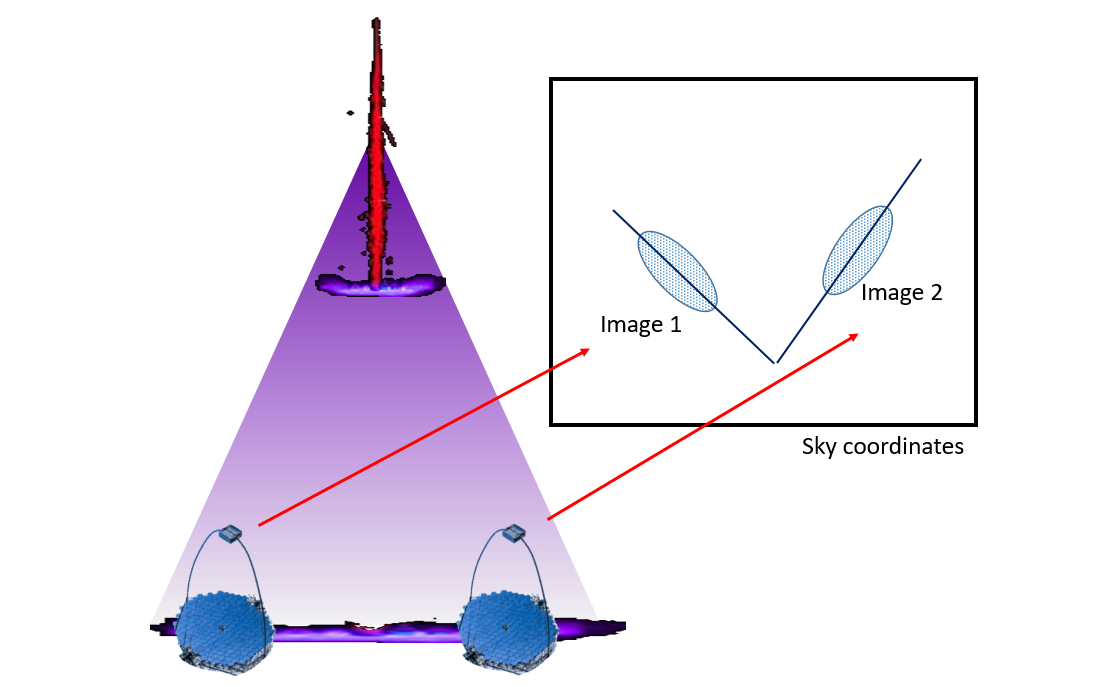}
    \caption{Two or more IACT record the Cherenkov photons (purple) emitted by the electrons and positrons (red) of a single electromagnetic shower. The right panel illustrates the projection of the recorded images into the sky.}
    \label{fig:stereoscopic}
\end{figure}

In addition, IACTs implement the stereoscopic imaging technique, illustrated in Fig. \ref{fig:stereoscopic}: two or more large convex reflector placed within the light pool, focus the Cherenkov light of a single shower onto the same number of cameras equipped with photodetectors. These cameras record the image of the shower from different perspectives, and the geometrical properties of these images allow to determine the properties of the primary particle. In particular, the crossing point of the longitudinal axis of these images projected in to the sky provides a determination of the direction of the primary, and the total recorded number of photons is directly linked with the energy. In order to accurately identify and reconstruct the primary particle properties it is necessary to make use of complex Monte Carlo simulations of the shower development and the detector response.

In order to implement this technique it is necessary to design a telescope with a very large optical aperture that allows to collect as many Cherenkov photons as possible, and a large FOV since the shower images can have an angular extend of up to few degrees and are shifted another few degrees from the source position in the sky. In addition, it is common to point slightly off-source to use a source-less portion of the sky within the FOV to estimate the background due to diffuse gamma rays and hadronic cosmic rays. These constraints lead to designs of the optics explained later in this chapter.

\begin{figure}
       \includegraphics[width=\textwidth]{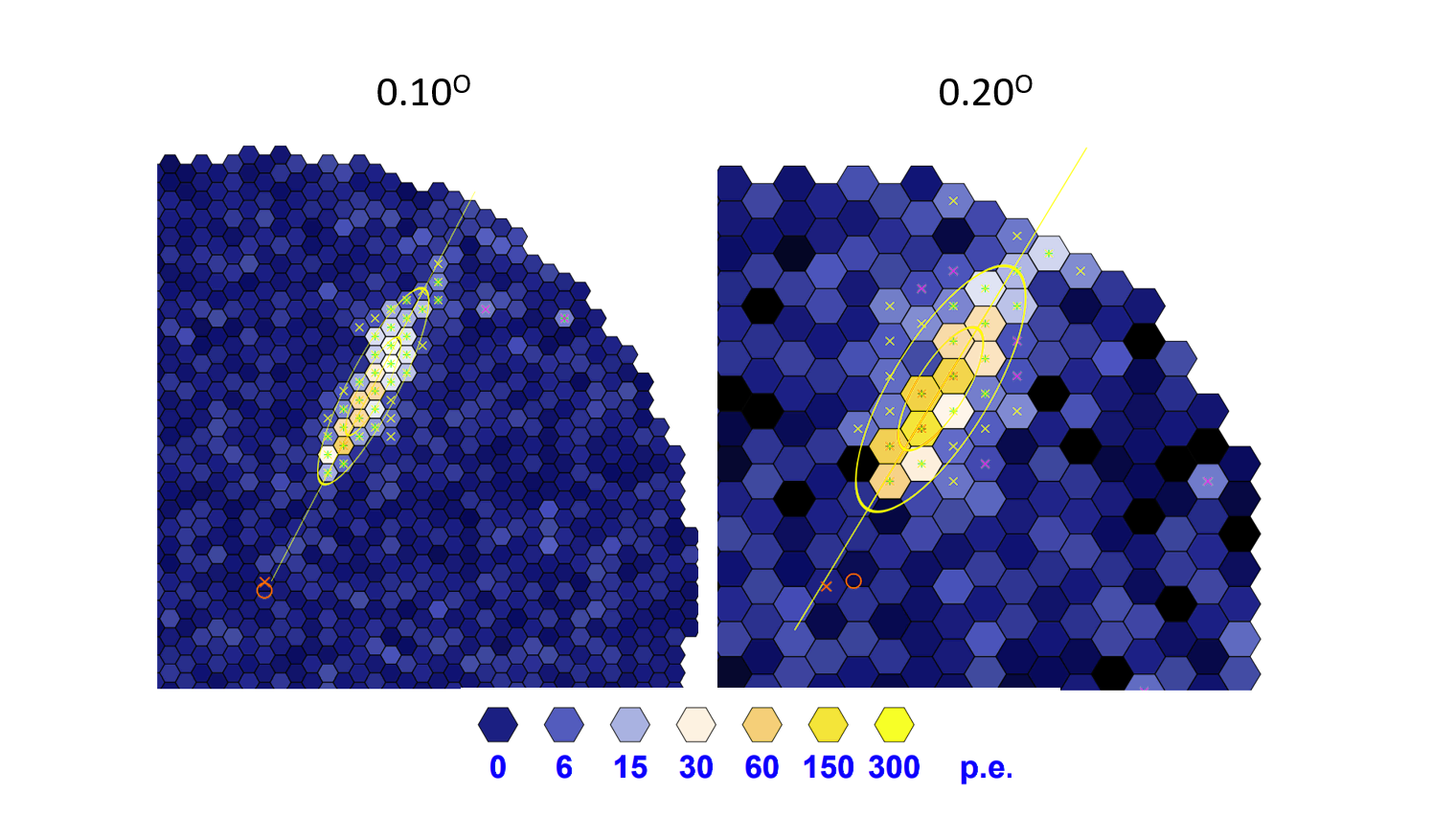}
       \caption{Simulation of the recorded image the same shower with two different pixel sizes.}
       \label{fig_pixel_size}
\end{figure}

To cover the required FOV cameras of 10-m type IACTs are usually large, with sizes measured in meters. Their focal plane is instrumented with fast detectors, with response times of order of nanosecond, high photodetection efficiencies (20\% or more), large detection areas, and very clean amplification, that allows to resolve single phe (phe) signals. These are described later in this chapter together with the associated electronics. The  size  of  focal  plane  pixels  is  a  parameter  which  requires  careful optimisation in IACTs. Figure \ref{fig_pixel_size} illustrates how a shower image is resolved at pixel sizes of $0.10^\circ$ and of $0.20^\circ$. The gain due to the use of small pixels depends strongly on the analysis technique.  In  the  classical  second-moment  analysis (Hillas analysis\cite{Hillas}),  performance  seems to saturate  for  pixels  smaller  than 0.15$^\circ$. On the other hand, analysis techniques which use the full image  distribution can  extract the information contained in the well collimated head part of high-intensity images,  as  compared  to  the  more  diffuse  tail,  and  benefit  from  smaller pixel sizes. Pixel  size  also  influences  trigger  strategies, since gamma-ray images are contiguous for large pixel sizes, allowing straight-forward topological triggers compared with the case with smaller pixel sizes. 

\section{Telescope optics}

Let us first consider the requirements for the telescope optics: 

\begin{itemize}
    \item As a general rule of thumb a $\gamma$-ray atmospheric shower produces 1 photon per m$^2$ and GeV. This means that one needs mirror collection areas of hundreds of m$^2$ to study $\gamma$-rays of $\leq$100~GeV energy. As a result, IACTs currently have the largest optical mirrors in the world. 

    \item However, as we have seen, shower images have an angular extent of tens of arcminutes and undergo intrinsic fluctuations due to the shower development in the atmosphere. Consequently, IACTs only require an optical point spread function (PSF) roughly of the size of a pixel, i.e. about 5~$\sim$arcminute in diameter. This contrasts sharply with the optical requirements of telescopes in the optical range of light, for which the PSF must be better than 1~arcsecond, and allows to relax the quality requirements for the mirrors and mirror support structure. 

    \item The FOV must have an angular diameter of at least 2~arcdeg to contain the images of low energy showers produced by point-like $\gamma$-ray sources and at least 4 arcdeg if the telescope is pointed slightly away from the source (''wobble mode'') and the intention is to observe high energy showers. Larger FOVs are needed if one plans to observe sources with an extent of a few arcmin or to perform sky surveys.

    \item As mentioned above, time is also of the essence in the identification and reconstruction of showers. The telescope optics must also satisfy a requirement for photon arrival time: the shape of the reflector must be isochronous to very few nanoseconds.

\end{itemize}

\begin{figure}
       \includegraphics[width=\textwidth]{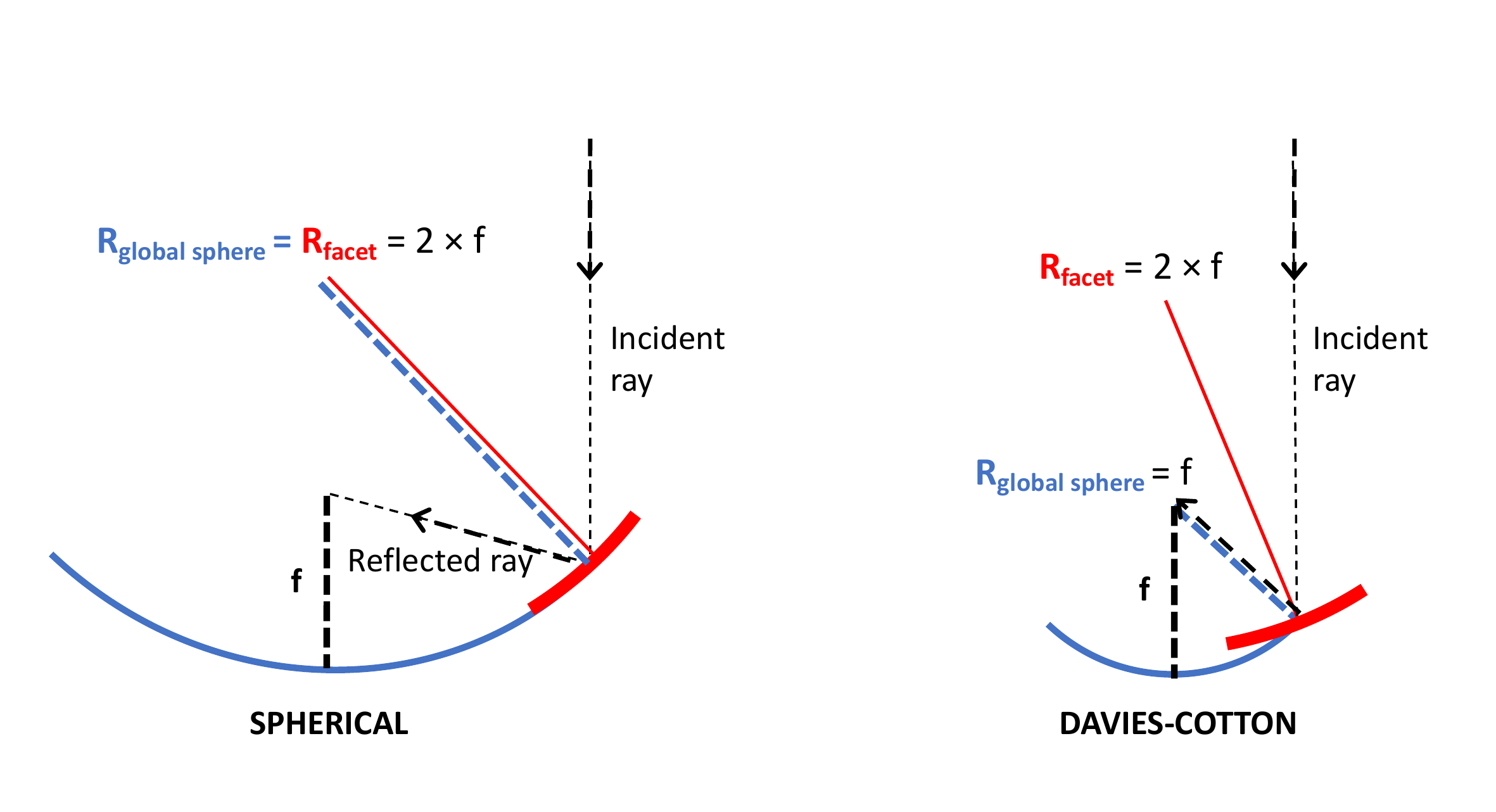}
       \caption{The left panel shows a spherical design: the radius of curvature $R_{sphere}$ is 2$\times f$ and the normal vectors of the individual facets point to the center of the sphere. The right panel shows a Davies-Cotton design. For the same $f$ the global shape of the reflector is spherical but has a smaller radius of curvature.
       At the same time the normal vectors of the individual facets are no longer aligned with the normal vector of the global sphere and their curvature does not follow its curvature.}
       \label{fig_davies_cotton_optics}
\end{figure}

Most IACTs follow a simple optical design. Light is collected by a convergent reflecting mirror surface (''reflector'') into a photodetector camera located at the focal plane. Like all large optical telescopes, IACT reflectors are multifaceted. Mirror facets have a typical surface area of 0.25 - 2 m$^2$. A spherical reflector shows too poor an optical performance. Two other reflector concepts are used: 

\begin{itemize}
    \item A parabola. The overall shape of the reflector is parabolic, while each facet has a spherical shape with a curvature corresponding to the local curvature of the parabola. All facets in a ring around the center of the reflector are equal. In principle, the individual facets should be aspherical, but in practice they are manufactured with the same radius of curvature in the sagittal and tangential directions. A parabolic reflector has an excellent PSF at the center of the FOV, but suffers from off-axis coma aberration. The main advantage of a parabolic is its excellent temporal resolution. 
    \item A ''Davies-Cotton'' design. Many IACTs have opted for the Davies-Cotton design\cite{Davies-Cotton}, which can only be applied when the reflector is multifaceted. Figure \ref{fig_davies_cotton_optics} shows how this design compares to a spherical reflector. For a given focal length $f$, the radius of curvature of a spherical reflector $R_{sphere}$ is 2$\times f$ and the normal vectors of the individual facets point to the center of the sphere. Instead, in a Davies-Cotton design, the reflector follows a global spherical shape with a smaller $R_{sphere} = f$ whereas the individual facets have a constant $R_{facet} = 2\times f$. However their normal vectors do not cross the center of the reflector sphere but a point at a distance $2\times f$ along the optical axis.
    
\end{itemize}

\begin{figure}
       \includegraphics[width=0.8\textwidth]{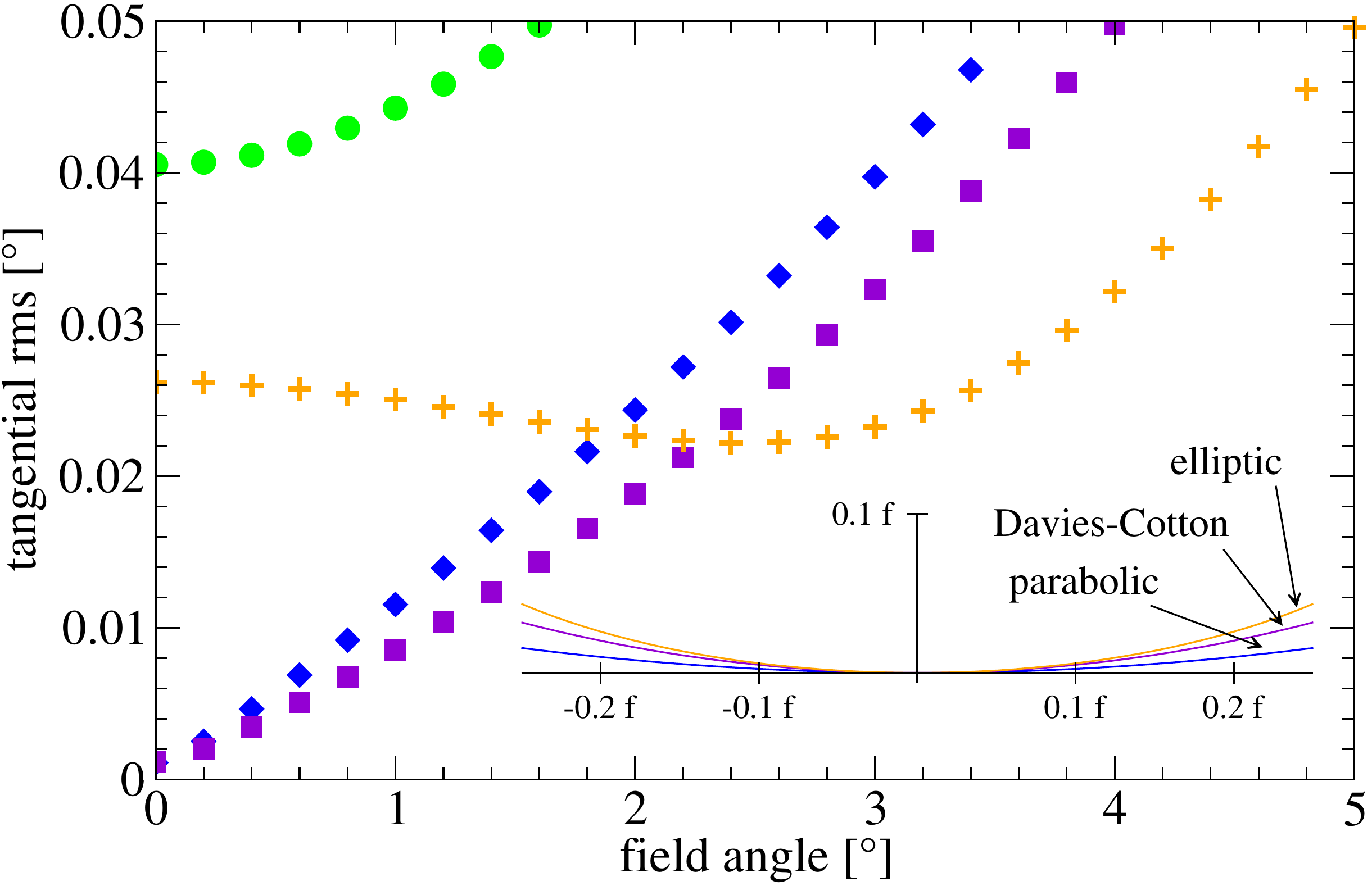}
       \caption{tangential RMS for given incident angle for a spherical design (green circles), a Davies–Cotton design (purple squares), a parabolic design with adjusted radii (blue rhomboids) and a more sophisticated elliptical design (orange crosses). The focal ratio is fixed to 2 and the tessellation ratio (ratio of area of facet and area of reflector) to 0.03. The inset shows the actual gross shape of the different configurations.}
       \label{fig_Schliesser-Mirzoyan}
\end{figure}

Figure \ref{fig_Schliesser-Mirzoyan} illustrates how the tangential RMS changes with incident angle for different optical designs (the sagittal RMS does not change with the incident angle). The results are based on a ray tracing simulation \cite{Schliesser-Mirzoyan}. For any of these optical designs increasing the ratio of focal length and reflector diameter (focal ratio, f/D) reduces aberrations but increases the cost and complexity of the telescope mechanics so IACTs do not exceed a focal ratio of 1.5.

Such a simple telescope optics has significant advantages: the design is simple, production is less expensive and there is no loss of light in the secondary optical elements. However some IACTs have opted for more complex optics in order to achieve a larger FOV and to reduce the plate scale. The latter allows to use smaller photosensors, namely Silicon PMs. The so-called Schwarzschild-Couder design\cite{Couder} has two aspheric mirrors that can be configured to correct for spherical and coma aberrations, achieving good optical quality over a large FOV of $\sim$10$^{\circ}$ with a small focal ratio and plate scale.

\subsection{Mechanical structure}
%
%
%
%

Compared to telescopes in the visible range where optical precision is higher, IACTs can be built with relatively simple mechanics and without a protecting dome. This reduces the cost considerably. 

However both the mirrors and the mechanical structure must be particularly resistant to the harsh atmospheric conditions common in astronomical observatories, more specifically to high ice loads or strong wind gusts. Working outdoors also requires low-maintenance technical solutions. 

In addition some phenomena in VHE astrophysics are very fast. In particular, the prompt emission of gamma-ray bursts lasts only a few seconds. In general, IACTs are expected to re-point in about 1-2 minutes but some IACTs have been designed to re-point to any position on the sky in a time as short as 20-30 seconds. Acceleration during rapid re-pointing introduces additional forces in the structure. 

Except for the initial smaller ($<$3 m) models, IACT are typically equipped with alt-azimuth mounts. 
The structure can be implemented as a positioner (tower, head and yokes) and a reflector support that rotates in azimuth and elevation, or alternatively as a space frame substructure that rotates in azimuth and a reflector support that rotates only in elevation. The reflector support can be implemented as a truss network or a space frame. The space frame elements are fabricated from steel or carbon fiber reinforced plastic (CFRP, crystalline carbon filaments inside a resin such as epoxy). 

CFRP is much stronger than steel in terms of strength to weight ratios (although an actual comparison should take into account the geometry of the carbon fiber filaments). That means that CFRP structures are significantly lighter than steel ones. 

The camera holds to the mechanical structure either through a steel truss structure or with a parabolic aluminium or CFRP arch in the vertical plane and additional tension cables. They all hold the camera to points at the edge of the optical support structure.

The mechanical accuracy of the structure must be significantly improved in dual-mirror telescopes. All implementations of dual-mirror telescopes attach the secondary mirror to the optical support structure with a lattice structure, and some of them add an additional support element closer to the optical axis. They are designed to achieve minimum shadowing, control stray light and, since the IACTs are not covered by a dome during the day, protect them from sunlight during daytime parking.

The mirror facets hold to the optical support structure normally through two positioning points and a floating one. The facet is aligned using the positioning support points, which are typically implemented as electric step motors (actuators) with a range of a few mm and steps of $\mu$m. Facets may be aligned at installation and every few months/years to correct for slow degradation in the mechanics (typically on rigid steel structures) or every 10-20 minutes to correct for changing gravity force, using an Active Mirror Control system (typically on less rigid CFRP structures). 

Active Mirror Control (AMC) calibration is based on a number of ancillary items. A CCD camera is installed in the center of the reflector pointing to the photodetector camera. A diffusing target can be placed near the focal plane. When the telescope is pointed at a bright star, the reflections of the star on the individual facets are identified on the target by the CCD camera. The reflections are referenced to LEDs installed on the edges of the camera. The AMC software aligns each individual facet until all star reflections fall at the center of the camera. This calibration is normally run for various elevation angles and the actuator positions are recorded and used during standard observations. The calibration must be repeated on time scales of years.

The same or a second camera is used to monitor the PSF during regular observations or a few times throughout the night to identify technical problems with the AMC. 
 
Other schemes for the AMC have been or are being tested, e.g., installing a laser or CCD camera on each facet.

Dual-mirror telescopes add the additional complication of aligning the secondary mirror independently of the primary. The accuracy of the secondary mirror is higher because it is located close to the camera. In general, this requires more degrees of freedom in facet alignment, implemented, for example, by Steward platforms.

\section{Mirror technology}

IACT reflectors are among the largest telescope reflectors in the world. That makes cost a significant driver in their design: one typically aim at a cost $<$ 2000 Euro/m$^2$. At the same time they must be as light as possible and resilient to environmental conditions. On the positive side optical quality is not as high as for telescopes in the visible range. Reflectivity should be as high as possible, typically exceeding 90\%, but mainly in the range from 300 to 500 nm where most of the Cherenkov light reaches the ground. Telescopes with dual optics pose extra difficulties due to smaller radii of curvature and aspherical shapes.

In general mirror facets have a high aspect ratio ($>$25), with a thickness of a 5-10~cm and 0.5 - 2~m side to side. Facets have either square or hexagonal shape. 

Mirrors are either ground from a raw blank and aluminized, or manufactured as a composite sandwich structure. Grinding is typically more expensive, takes longer to produce and is heavier, so composite mirrors are more and more in use. A sandwich mirror is a structure with a face plate, a light weight honeycomb core providing stiffness, and a base plate anchored to the telescope structure. The face plate may be machined to the desired curvature, glued to the core and machined. Alternatively, a thin glass sheet of a few mm with its reflective surface may be pressed against a mold using vacuum suction to reach the required curvature (``cold slumping''). The sheet is later glued to the honeycomb core.

The reflective layer in the face plate was traditionally evaporated aluminium although dielectric layers are also used nowadays (e.g. HfO$_2$, TiO$_2$, SiO$_2$). An advantage of a multi-layer coating is that the reflectivity spectral response can be adjusted, for instance to reject night sky background at longer wavelengths. The reflective layer is applied on top of a thicker layer of glass or metal and is typically protected from the environment (mechanically and chemically) using an external layer of $\sim$100 nm of quartz.

\section{Telescope control, event reconstruction and data products}

Similar to other telescopes, the control system of an IACT has to deal with a number of subsystems, namely the data acquisition (DAQ), trigger, telescope drive, camera electronics and mechanics control, mirror alignment control, starguider and other auxiliary subsystems (power, absolute clock, condition monitoring, atmospheric monitoring etc). Response times in the control system are typically of a few ms. Most of the IACTs operate in an array so a central array software is generally implemented on top of the individual telescope control.

An accuracy of target pointing and tracking better than a few tens of arcseconds is a usual requirement. 
As a first step the pointing position is monitored using angular encoders on both telescope axes, which ensure real-time that there are no significant tracking errors. 
But, to achieve the required precision, the telescope needs a pointing model that typically involves the drive and, in the case of IACTs with AMC, also the facet alignment model. The pointing position is measured using a starguider, implemented as a CCD at the center of the telescope reflector that registers simultaneously the position of the camera and the position of a bright stars in the sky. Typically $\gamma$-ray data are corrected offline for deviations in time scales of seconds to minutes using starguider data whereas the telescope pointing model is updated only on time scales of months.

The response of the camera photodetectors and front-end electronics is monitored using a light calibration source 
that flashes the camera uniformly. This calibrator generates fast pulses mimicking Cherenkov light pulses in time profile and brightness and is typically implemented with a laser. Calibration pulses may be interleaved with air shower pulses. Random triggers ("pedestal events") are recorded as well to determine the level of background noise.

The telescope subsystems are typically monitored in time scales of 1-10 seconds and some key parameters that are necessary for the $\gamma$-ray processing enter a "control data stream". 

Air shower, calibration and pedestal events constitute a separate high-throughput ($>$GByte/s) "raw data stream". Raw events are delivered by the DAQ and contain information from each pixel. The Cherenkov pulse is typically digitized at the front-end electronics with sampling rates up to 2 GSample/s. The pixel information included in a raw event may vary from just the total integrated charge and one single arrival time, to the whole digitized waveform over a period of a few tens of ns. In addition the event includes an event number, a time tag with a precision of at least a few hundred ns, a tag with information of trigger type (shower, calibration, pedestal, stereo, single-telescope etc) and other control and auxiliary tags.

\section{Photosensors}
%
%
%
%
%
%

The photosensors most commonly used in IACTs are photomultipliers with alkaline photocathodes and dynode chain-based electron multipliers, which provide ultrafast signals and allow measuring single phe. They can reach a relatively high peak quantum efficiency (QE), of about 40\%, with a dynamic range of about 5,000 phe. However these are not the only possible photosensors that can be employed in IACTs and new devices, like solid-state Silicon Photomultipliers (SiPM), are becoming viable candidates for new telescopes.

Generally speaking, any device is a viable photosensor if it fullfils the following criteria:
\begin{itemize}
    \item The photosensor should allow to determine the arrival time of the photons to each pixel with sub-nanosecond precision for light pulses of sufficient large amplitude. This is necessary to avoid distorting the intrinsic time evolution of the recorded shower. 
    \item The Cherenkov spectrum peaks at 350~nm and has a cutoff below 300~nm, therefore the peak of the efficiency of the photosensor must match this wavelength range. In addition, given that the NSB contribution increases with the wavelength, it is desirable that the sensor efficiency drops to zero above 550nm\footnote{This is not completely correct: there are re-emission lines of Cherenkov radiation above 650~nm due to the rotation states of OH molecules.} to reduce the background.
    \item The dynamic range of the sensor must be broad enough to accurately reconstruct showers initiated by gammas across a wide range of energies. Typically, a range that starts at one phe and extends to a few thousand phe provides a good balance between cost and performance, allowing for the reconstruction of showers over a range of three orders of magnitude in energy without the need for extrapolation of truncated signals.
    \commentout{The dynamic range of the sensors must cover from one phe to few thousands of them, with good linearity (few percent) or with a non-linearity that can be corrected up to this level. This is necessary to detect and reconstruct showers initiate by gammas within an energy range as wide as possible.\comment{Mazin: I don’t see how a dynamic range of few thousands results in an energy range “as wide as possible”. Why not few hundred or 100 thousand? I would rather argue that a dynamic range of few thousands provides a good compromise of cost versus performance, i.e. 3 orders in energy range, which can be reconstructed easily (no extrapolation of truncated signals) }}
    \commentout{\item Cross talk between adjacent sensors must be below the fluctuation of the signal in the whole dynamic range. Therefore it must be below the 1\% level.\comment{This is not true for SiPMs. One can leave with 10-20\% cross talk, simply the resolution suffers accordingly.}}
    \item The contribution of spurious signal to the trigger thresholds, and to the trigger rate, shall be negligible. This is specially important for photomultipliers, in which ionized atoms trapped in its interior can give rise to large signals long after true phe have been identified. For SiPM a similar phenomenon takes place, the optical crosstalk, although the mechanism is completely different\commentout{\cite{ref: optical cross talk}}. The rate of spurious signal is typically required to be at the $10^{-4}$ level or below with respect to the true signal. 
    \item In order to reduce the variance of the signal, the uniformity of the response within a single sensor must be better than 10\%. This requirement can be extended to the uniformity between different sensors in the same camera.
    \item Given that the sensors can be exposed to indirect sunlight during maintenance operations, or to indirect moonlight during observations, they have to be able to survive to strong illumination. In particular, this survival requirement during observations  imposes the use of a current-limiter in the high voltage supplier for photomultipliers, whereas SiPM do not require such protection.
    \item The size of the sensor should match the typical angular size of the fluctuation of the shower images in the camera focal plane, and should not introduce dead areas in the camera. This cannot be usually fulfilled by off-the-shelf sensors. However it can be achieved by coupling a light guide\commentout{Winston cone\comment{MAzin:Maybe better “light guide”? Probably a picture would help here as describing typical light guide in words is not trivial.}} to it, designed to have a pixel FOV of about 0.1 degrees. Moreover, this design allows to reject most of the background light outside this FOV and to keep the dead area of the camera small, without having to used custom geometries for the sensors.
\end{itemize}
Apart of these criteria, there are other operational aspects that have to be considered when selecting photosensors for IACTs. In particular  their lifetime, stability and cost are important aspects to consider in order to keep the performance of the camera up to the expectations, with a bounded maintenance cost.

\section{Camera Trigger and DAQ}
Triggering in telescopes have two goals: to reduce the load of the data acquisition system to manageable levels, and to reject background events, mainly due to NSB fluctuations. IACTs are usually triggered in a staged manner. Firstly, individual Cherenkov telescope cameras produce a {\bf camera trigger} \commentout{Mazin: The camera trigger is typically a staged trigger as well} signal by making use of topological properties in the pixel signals. This camera trigger may be staged itself and is designed to minimize the probability of triggering due to fluctuations of the NSB. The signal is either sent to a centralized facility or to neighbouring telescopes to build a higher level trigger signal by exploiting the temporal coincidence of camera trigger images within the array of telescopes, the so-called {\bf stereo trigger} or {\bf array trigger}. The array trigger is sent back to the cameras to proceed recording the images. Therefore an individual telescope typically buffers the shower image from the moment the camera trigger is issued to the moment it receives the stereo trigger. This can be done without introducing dead times in scales of ms for digital buffering, or $\mu$s for analog buffering. Therefore, the kind of buffer has a strong influence in the trigger design. Regardless of the chosen implementation, the trigger system must be flexible and software-configurable, since operation modes vary from deep observations, where all telescopes follow the same source, to monitoring or survey applications, where groups of a few telescopes or even single telescopes point in different directions.

\subsection{Camera trigger}
The {\bf camera trigger} must keep the trigger rate due to fluctuations of the NSB low. For this purpose, it exploits the recognition of the pattern due to the concentrations of Cherenkov signals in local regions of the camera. Nowadays this recognition is based in looking for a number of pixels above threshold, or a number of neighbouring pixels above threshold, within the camera. This is typically implemented by dividing the camera up into sectors, which must overlap to provide a uniform trigger efficiency across the camera. In an alternative approach the sum of all pixels signals in a patch of neighbouring pixels, capped to a maximum value to reduce the influence of afterpulsing,  is formed, and a threshold is set to initiate a trigger. In both cases the implementation can be digital or analogue, although the second one is usually employed. The decision signal is sent in digital format, sometimes together with additional timing information, to neighbouring telescopes or to a central decision system to form the stereo trigger.

\subsection{Stereo trigger}
The stereo trigger schemes for systems of IACTs provide asynchronous trigger decisions, delaying individual telescope trigger signals by an appropriate amount to compensate for the time differences when the Cherenkov light reaches the telescopes, and scanning trigger signals for patterns of telescope coincidence. The time to reach a trigger decision and to propagate it back to neighbouring telescopes is of the order of $\mu$s. This can raise up to the ms scale when the signals have to be propagated to a central unit to take the decision.
If the data is digitised and buffered with the individual camera trigger or with a lower level stereo trigger that combine individual triggers of neighbouring telescope, restrictions on array trigger latency of a centralized decision are greatly relaxed, and the decision can be software based. In this scheme, along with each local trigger an absolute timestamp with an accuracy of 1 ns is sent to the central decision system, which searches for time coincidences of the events and defines the telescope system trigger. The centralized trigger system sends the information of the coincidence time back to the cameras. The cameras then select the events that fulfil the global trigger condition and should be recorded and transmitted for further stereoscopic processing. This centralized scheme is software-based, but still makes uses of the properties of the individual camera trigger system in an optimal way. 

\subsection{DAQ electronics}
As already commented, shower images have a pulse width of a few ns, with a background due to the NSB with typical rates from tens to hundred of MHz per pixel depending on mirror and pixel size and the photodetection efficiency. Thus, recording the shower induced signals efficiently requires high bandwidth and short integration times. On top of that, the dynamic range and electronics noise should be such that \commentout{few phe signals are resolved \commentout{Mazin:As far as I know, in several experiments the single phe are not resolved (example MAGIC). Resolving the single phe helps the calibration but there are different ways to do it, e.g. F-factor method for PMTs}, and} signals from one phe up to at least few thousands are recorded without truncation. Given the latency of the trigger signal, the electronics must delay or buffer the signals until the decision to store the signal arrives, which can take up to about 10 $\mu$s if the trigger signal of several telescopes are combined. Current available signal recording and processing technologies allow recording a range of signal parameters, from the integrated charge to the full pulse shape, over a fixed time window. The latter option seems optimal for low energies (below few TeV), that requires a  more sophisticated background reduction, whereas for higher energies the former parameter is usually enough, complemented with other few parameters like time and time width of the signal.

Two techniques for signal recording and processing are in use nowadays. The first technique is based in the use of Flash Analogue-to-Digital Converters (FADCs), while the second one uses analogue sampling memories:
\begin{itemize}
    \item{FADC technology:} these digitise the photosensor signal at sampling rates between 100 Megasamples/s and few Gigasamples/s. The digitised stream can be subsequently stored digitally for further processing, which allows for longer trigger latencies. Moreover, this technology allows the implementation of fully digital trigger systems that exploit the recorded image in real time. The main disadvantages of using FADCs with respect to analogue sampling memories are their cost and power consumption, which makes difficult its integration in cameras, although recent developments on low-power and low-cost FADCs with sampling speeds of up to 300~MS/s,  make them competitive for some IACT cameras.
    \item {Analogue sampling memories technology:} these consist of banks of switched capacitors which are used in turn to record the signal shape. The maximum recording depth is given by the number of capacitors and the sampling time.  The implementation in ASICs is such that they allocate enough capacitors to cope with few microseconds of trigger latency at sampling speeds 1~GS/s or faster for several channels simultaneously, making them very competitive in terms of cost and power scalability, and allowing a full implementation inside the camera.
\end{itemize}

Once the event signals have been sampled and digitised, they can be processed either in FPGAs to perform a first reduction of the information by storing only pixel charges and pulse time width,  or they can be fully stored for subsequent offline processing. In either case, the transmission system and consumer electronics and software must be able to deal with trigger rates of up to 20 kHz per camera, in which the signal of few thousand pixels, and few tens of time samples, if not reduction is performed, has to be buffered and eventually transmitted for archival. Currently this can be implemented using local digital buffers in the cameras, and commercial hardware for transmitting the data from the camera to high-end computers, which can assemble the events in real-time, and store them on disk.

%
%
%
%
%
%
%

\section{Analysis techniques}


%
%
%
%
%

The analysis of gamma rays detected using IACTs has to cope with different sources of background at different stages. At the earliest one \commentout{Mazin: As far as I know, in several experiments the single phe are not resolved (example MAGIC). Resolving the single phe helps the calibration but there are different ways to do it, e.g. F-factor method for PMTs}, when the signal collected on individual pixels is integrated to obtain the number of phe, the integration has to be performed in such a manner as to reduce the contribution of electronic noise and the fluctuations of the ambient random photon field, which is mainly due to NSB and diffuse Moon Light. At later stages, the recorded shower images have to be treated to minimize further the contribution of the fluctuations of that random photon field. The techniques used in these two first stages are called Signal Extraction and Image Cleaning respectively. Finally, at the latest stage of the analysis, it is necessary to reject the population of recorded images that are not initiated by gamma rays from the object of interest for the data analyzis. In doing so the technique to be used depends on the physical origin of the image, which is mostly showers initiated by hadrons and gammas from the object source, but could also include Cherenkov rings due to atmospheric muons, diffuse gammas or electrons.  Let us discuss the rejection techniques and the main properties of the backgrounds.

\subsection{Signal extraction}
\commentout{Mazin:This is a bit too detailed for my taste but up to you. One aspect which I did not find here is a possibility to use information from neighboring pixels to predict expected arrival time for Cherenkov photons which largely reduces contribution from NSB }
The NSB light level near the zenith at typical locations of IACTs is about $2.4\cdot10^{3}$ photons sr$^{-1}$ ns$^{-1}$ m$^{-2}$ for wavelengths between $300$ and $650$ nm \cite{Preuss et al.}. For typical IACTs this results in a time average detection level of about 0.1 phe ns$^{-1}$ per pixel, which can increase up to $10$-fold in the presence of diffuse Moon Light. \commentout{Given the usual number of pixels, which is about few thousands, the total number of background phe is about few hundreds per nanosecond, with is of the same magnitude order as the total number of phe spread in few nanoseconds for showers initiated by gamma rays with an energy of 0.1 TeV.}
On the other hand, the total number of phe detected with a typical IACT in a gamma ray with an energy of 0.1 TeV is of the order of few hundreds spread over few nanoseconds, scattered in about 10 pixels\commentout{\cite{signal for IACTs}}\footnote{For pixels of a typical size of 0.1$^{\circ}$ and a 10-m type IACT}. 
The goal of the signal extraction algorithm is to identify the time of arrival and number of phe in each pixel within a recorder window that lasts few tens of nanoseconds. It profits from the clustering in time of the phe due to showers reaching a given pixel. To this end it has to identify the most probable location in time of the signal on each pixel, and integrate it in a time window large enough to capture it, but small enough to minimise the amount of noise accounted. This width relates to the bandwidth of the system, that usually dominates the time width of the signal recorded for a single phe. 
Most signal extraction algorithms rely on minimizing the following goodness of fit $Gof$ between the recorded signal of a pixel in a given time window and the expected response function:

\begin{equation}
    Gof=\sum_{i\in n} (S_i - N~R_{i,T}))^2=\underbrace{(\sum_i S_i^2)}_{constant} +N^2~\underbrace{(\sum_i R^2_{i,T})}_{constant} -2\times N\times\underbrace{(\sum_i S_i R_{i,T})}_{correlation~between~R~and~S}
    \label{eq:signal extraction}
\end{equation}

where the summation is in the $n$ samples in time, $S_i$ is the signal in sample $i$, $R_{i,T}$ is a given normalized response function centered in time sample $T$ and evaluated in sample $i$, and $N$ is a normalization factor. The rightmost hand side of the equation above separates the $Gof$ in the terms which are relevant for the discussion of the signal extraction. The minimization of $Gof$ is usually with respect to $N$ and $T$, which gives a time and amplitude for the signal. Depending on the response function $R$ and how the minimization is performed the usual strategies are:

\begin{itemize}
    \item {\bf Fixed window}: $T$ is kept fixed, $R$ is constant in a window around this sample and zero out of it, that is, a square function of a given width. With this $N$ is proportional to the average of the signal within this window.
    \item {\bf Sliding window}: $R$ has the same shape as for the fixed window, but $Gof$ is minimized with respect to $T$. In this case $T$ is the one that maximizes the correlation between the signal and the square function $R$, and $N$ is the proportional to the average around this $T$.
    \item {\bf Digital filter}\cite{digital filter}: $R_{i,T}$ has the expected shape for a single phe arriving at $T$. The minimization gives the value of $T$ that maximizes the correlation between the signal and $R$, and $N$ is proportional to the weighted average of the signal around the sample $T$ with the chosen response.
\end{itemize}
The main drawback of these methods relies in the assumption of the response function shape in eq. \ref{eq:signal extraction}. Selecting the right one is a key parameter to reduce the influence of the electronic noise in the extracted signal and noise. On top of that, independently on the chosen response function, variations of the shape of the true signal result in a fluctuation in both, the reconstructed time $T$ and the extracted signal normalization $N$, which are undesirable.

These are not the only possible strategies, and there are algorithms in the literature which aim to improve the background rejection and to reduce the systematic error in the value of the extracted signal, like using smooth basis functions \commentout{\cite{splines extraction}} to describe $R$, using the information from neighboring pixels
to predict the expected arrival times of the Cherenkov photons, or employing state of the art techniques like deep-learning methods. \commentout{\cite{deepl signal extraction}} \commentout{Nevertheless, the impact of these improvements is shadowed by the performance of the image cleaning procedure, which is the next step in the background rejection, except for methods based in whole image processing like some deep learning ones.}
Except for the methods based in whole image processing, like some deep learning ones, the result of the image cleaning algorithm, i.e. a list of arrival time and total signal for each signal, is used as input for the next step in the analysis: the image cleaning.

\subsection{Image cleaning}
\commentout{Mazin:It might be important to mention that there is a possibility to avoid image cleaning either completely or relax it significantly by comparing every calibrated event with some model which is typically a combination of what is to be expected from a gamma-ray shower + measured noise from the data. Carlos: actually this is said at the end of the section already. I will stress it.}
The task of an image cleaning procedure is to identify as many pixels as possible which are dominated by the signal from the shower, rejecting the pixels dominated by noise. To this end, the method exploits the correlation of the signal due to a single shower between neighboring pixels, which is not present for the background noise.

To understand how these algorithm works, it is useful to introduce a simple model that describes the mean instantaneous signal at time $t$ of a pixel with coordinates $x,y$ with respect to the maximum of a recorded electromagnetic shower. This simple model is based in the Hillas parameterisation\cite{Hillas} of shower images, approximating the time dependence of the arrival time of photons to the telescope by a linear function along the axis of the shower images. With this we have that the recorded signal $S(x,y,t)$ is

\begin{equation}
    S(x,y,t)= \big( b + A e^{-\frac{1}{2} \big( \frac{x^2}{w^2} + \frac{y^2}{l^2}\big)}\times e^{-\frac{1}{2}\frac{(y-vt)^2}{\sigma_t^2}} \big) \Delta
\label{eq: shower development}
\end{equation}

where the shower axis moves along the Y axis, the shower image maximum amplitude is $A$, $w$ is its width and $l$ its length,  $\sigma_t$ is the typical time length of a single phe pulse, $v$ gives the shower image development speed, $b$ gives the background signal per unit time, and  $\Delta$ is the sampling time width. The values of $A$, $w$, $l$ and $v$ depend on the shower characteristics (like gamma or hadron energy, impact point of the shower axis on the ground, ...) as well as the optics of the telescope, camera pixel size and detection efficiency. On the other hand $\Delta$ and $\sigma_t$ are mostly due to instrumental characteristics like the sampling speed and the full chain electronics bandwidth.

The signal extraction algorithm previously described extracts the time for each pixel in which eq. \ref{eq: shower development} is maximized as a function of $t$, which call $S_{extracted}(x,y)$. For pixels that record a large number of phe due to the shower compared with the NSB contribution, the resulting amplitude of the signal extraction will be $$S_{extracted}^{shower}(x,y)=Ae^{-\frac{1}{2} \big( \frac{x^2}{w^2} + \frac{y^2}{l^2}\big)}\Delta$$, and the time will be given by $y/v$.

On the other hand, for pixels that only contain background phe, the average number of recorded phe would be $b\times \Delta \times N$, where $N$ is the total number of time samples per pixel used as input for the signal extraction algorithms. These photons will be uniformly distributed among all the $N$ samples, and if $b\times \Delta \times N<\frac{\Delta}{2\sigma_t}N$, they will likely not overlap. Under these usually realistic conditions, the extraction algorithm will result in the maximum amplitude for a single phe $S_{NSB}$, and the time will be random among all samples. 

A first approach to reject pixels dominated by noise is to select only those such  that extracted signal is larger than $S_{NSB}+F$, where $F$ is a safety factor to account for uncertainties in the response, electronic noise and statistical fluctuations. However this is not fully satisfactory because requires to adjust the rejection level without any a priori knowledge of the shower image parameters. Since the required rejection efficiency will depend on the number of pixels, this results in either keeping NSB-dominated background as part of the image, or rejecting pixels of the shower image. Both possibilities result in introducing a systematic error for showers produced by low energy gammas or by high energy gammas far away from the telescope, and in the rejection of full images in case of showers with a small, but significant, number of photon-electrons detected.

\commentout{consider that the number NSB photons in a typical time window of few tens of nanoseconds containing the shower will be small. If this time window is $\Delta T$, the worst case for the NSB is all photons reaching a single time sample, that results in a maximum average number of phe of $b\times\Delta T$. Therefore, it is possible to reject pixels dominated by the NSB by selecting only those with $S_{extracted} > b\times\Delta T + C \sqrt{b\times\Delta T}$, where $C$ provides the rejection level. 
}

An improved approach is built by realizing pixels dominated by the shower cluster around the image maximum, so given a pixel which is safely considered not due to the background, its next-neighbours are likely due  to shower photons too. This idea is implemented by using two background rejection levels instead of one. The first one is tight enough to select a list of pixels which are due to shower photons, typically selecting pixels with a signal larger than the one expected for 10 phe. Then other pixels are added recursively to this list if they are neighbours of pixels already contained in the list and other signals are above a second looser rejection level, which usually are close to 5~phe. To improve the performance of the selection, as well as its robustness against background signals that do not fulfill the conditions for the background considered above, some additional constraints are required to add pixels to the list: a) pixel above the first level are only added if they have at least a next-neighbor pixel also above this level (these are called core pixels); b) pixels which are above the second but not the first level are only added if at least one of its next-neighbours is above the first level (these are called boundary pixels).
An additional improvement to this consists in taking into account the extracted time $t$ between pixels\cite{image cleaning}, or performing a global fit in time and space to a model similar to that of eq. \ref{eq: shower development} \cite{likelihood approach}, thus increasing the signal over noise ratio of the reconstructed shower image. Employing sophisticated models, this latest approach allows to relax the necessity of the event cleaning or even eliminate it completely.

 \commentout{Some plots about performance (from MAGIC for example?)}


\subsection{Gamma-Hadron separation}

\begin{figure}
    \centering
    \includegraphics[width=\textwidth]{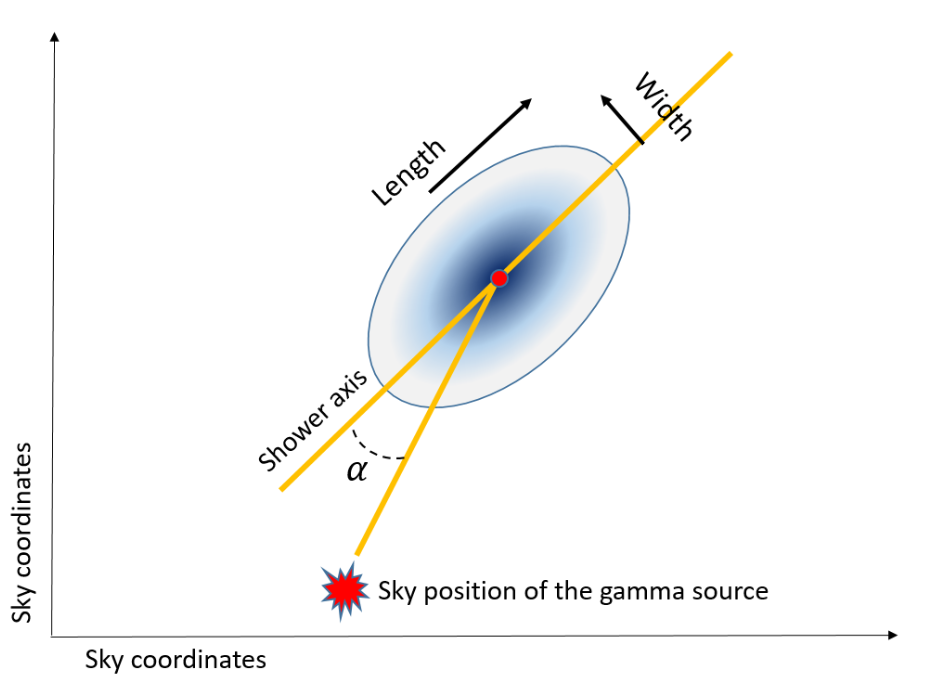}
    \caption{\commentout{Mazin:Should Length and Width be half-length and half-width?} Sketch of some of the Hillas parameters obtained by fitting the image of a shower recorded in a single telescope.}
    \label{fig: hillas parameters}
\end{figure}

The above-mentioned techniques deal mainly with the NSB. However, once the images have been cleaned, the main background are shower images produced by the interaction of non-gamma primaries with the atmosphere. These are dominated by protons and He nuclei, that constitute more than 95\% of primaries. The traditional rejection method for this background is based in the aforementioned Hillas parameterisation \cite{Hillas} and the exploitation of the stereoscopic observation of the showers. This consists in fitting the distribution of photons in the cleaned images of the same shower observed by all Cherenkov Telescopes to  bivariate gaussian distributions (one per image). The result of this parameterisation are total amount of light contained in the shower, the position of the core of the shower in the sky, the projection of the axis of the shower in the sky, and the length of the major and minor axes of the shower, shown in \ref{fig: hillas parameters}.

\begin{figure}
    \centering
    \includegraphics[width=\textwidth]{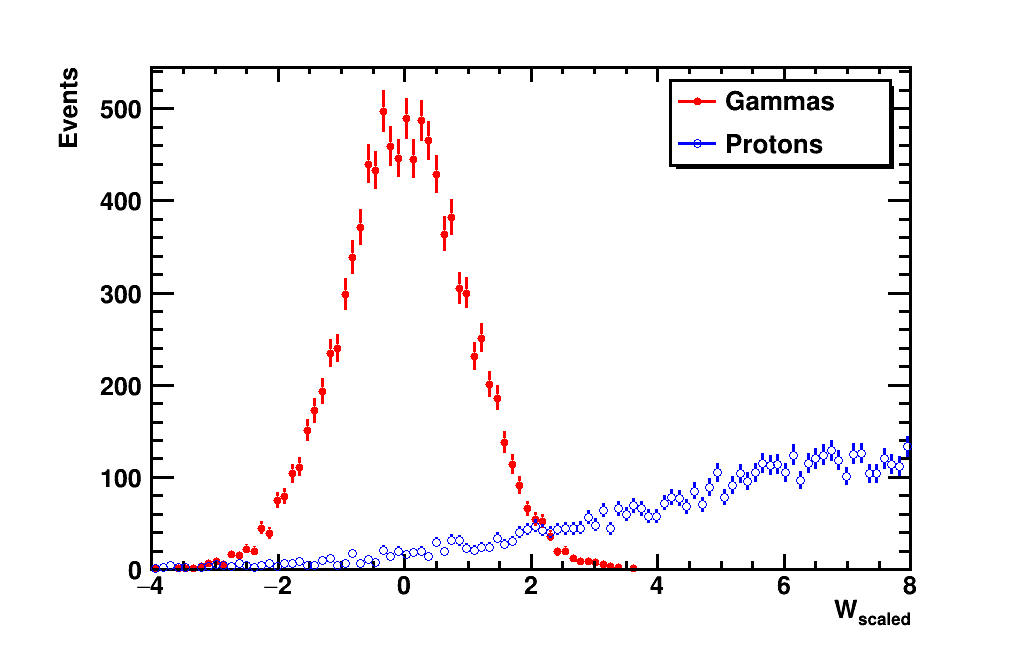}
    \caption{Distribution of $W_{scaled}$ for 4 telescopes and 1 TeV gammas and protons, obtained from a Monte Carlo simulation.}
    \label{fig:width}
\end{figure}

Since the image of an electromagnetic shower is well described by a single compact distribution except for very low energy events, a first level of rejection is traditionally obtained by discarding events in which at least one telescope recorded an image that contains more separated shower. 
On top of that, given that the shower axis follows the original direction of the gamma, a second level is achieved by rejecting events in which the projected shower axis of the parameterisation of all the recorded images of the same shower do not cross, within the expected resolution obtained from Monte Carlo simulation, in the same point in the sky. 
A final rejection level is the rejection of non-electromagnetic initiated events. In traditional analyses it exploited the Width parameter, which is directly linked with the Molière radius in the upper atmosphere, and, contrary to other Hillas parameters, has a weak dependence on direction or energy of the primary gamma. A discriminating variable can be built using the Width obtained from each telescope scaled by adding them in quadrature scaled by the expected one obtained from Monte Carlo simulations as a function of the cleaned image amount of light  on each telescope, and the distance of the telescope to the extrapolated impact point of the shower direction on the ground (the so called impact parameter).

\begin{equation}
    W_{scaled} = \frac{1}{\sqrt{N_{telescope}}} \sum_{i\in images} \frac{Width_i - <Width_i>(Size,D)}{\sigma_i(Size,D)}
\end{equation}

where $Size$ is the measured amount of light and $D$ is the impact parameter, $<Width_i>(Size,D)$ is the expected Width for telescope $i$, and $\sigma_i(Size,D)$ is the expected variance of the Width. Figure \ref{fig:width} shows the distribution of the $W_{scaled}$ for gammas of 1 TeV and the proton background for observations with 4 telescopes and gammas, showing that a simple cut can reject most of the proton initiated showers.

More sophisticated methods make use of all Hillas parameters simultaneously using multivariate classificators like Random Forests or Boosted decision trees \cite{random forest}. There are also extension of the Hillas parameterisation, for example the use of templates in the so-called Model Analysis \cite{model analysis} or fitting all the images to a 3D model for the shower \cite{3d reconstruction}, or lately using Deep neutral network models \cite{dl}. These provide an improved background rejection efficiency over the traditional one, at the cost of a larger complexity. They are especially useful to enlarge the lower end of the energy range covered by the telescope, where the statistical fluctuations of the images make the classification based on Hillas parameters less performing.

\commentout{threshold of the telescope trigger system the proton background is usual larger due to its power law distribution being softer than that of a gamma-ray source \commentout{MAzin:also gammas have typically a power law distribution. Maybe something like “because the energy distribution of the hydronic background has typically a softer energy distribution than the one of the gamma-ray source” }, and the recorded shower images have larger statistical fluctuation which makes more difficult the classification. \comment{Mazin:Makes the classification based on Hillas parameters less performant?}   }

\subsection{Determination of gamma-ray energy and incident direction}

The final stage of the analysis of a sample of shower images taken with IACTs, once backgrounds have been rejected, is the determination of primary gamma-ray direction and energy on an image by image basis.

\begin{figure}
    \centering
    \includegraphics[width=0.8\textwidth]{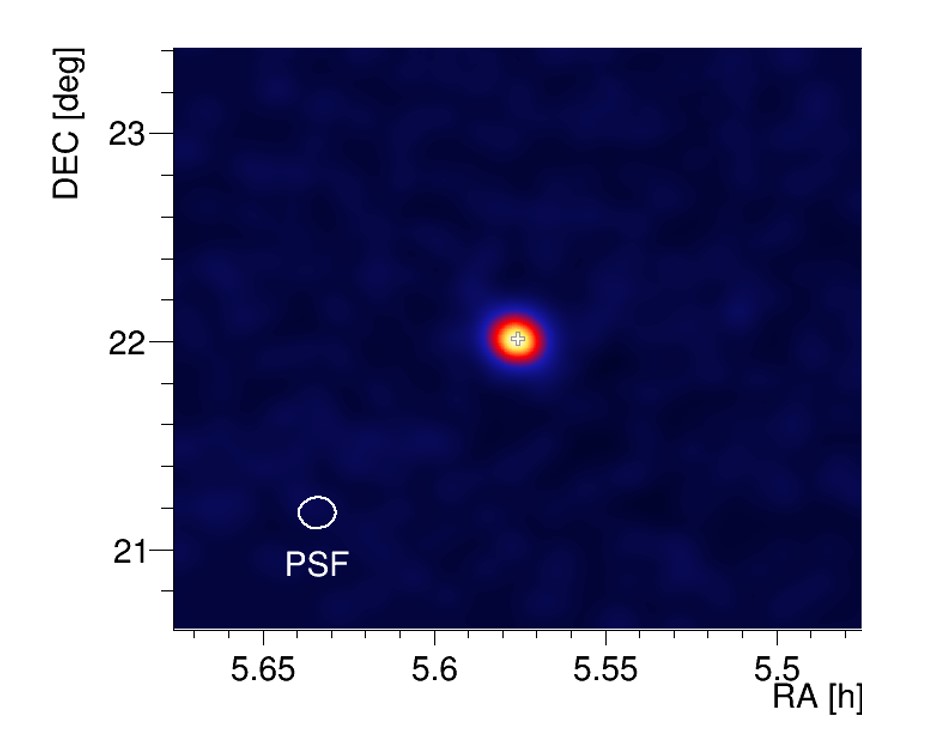}
    \caption{Distribution of the reconstructed direction of the identified gammas projected on the sky directions for a typical VHE gamma-ray source measured by a typical Cherenkov telescope. The color scale reflects the number of background subtracted gammas detected, smoothed to reduce the visual effect of statistics fluctuations.}
    \label{fig:skymap}
\end{figure}

\begin{figure}
    \centering
    \includegraphics[width=0.8\textwidth]{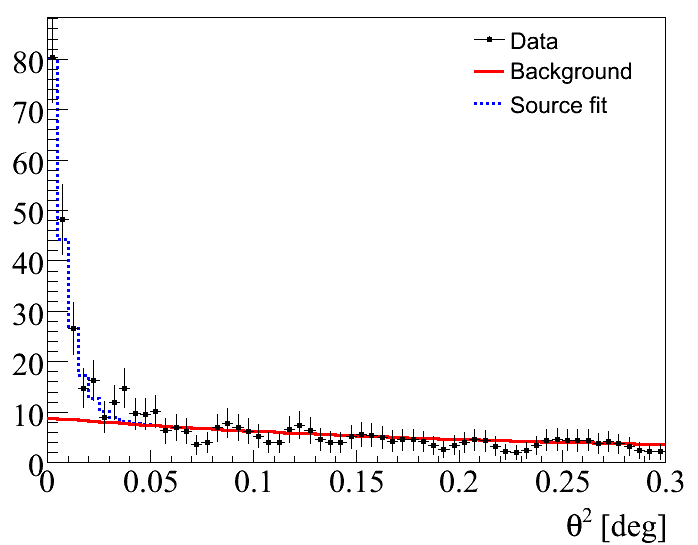}
    \caption{Distribution of the square of the angular distance of the inferred incoming direction of the $\gamma$-ray showers with respect to the known source position (black points) for the source of Fig. \ref{fig:skymap}, superimposed with the estimated irreducible background (red solid line), and a fit to a point like source (dashed blue line).}
    \label{fig:theta2}
\end{figure}

The accurate determination of the direction is possible thanks to the stereoscopic observation of the shower, as already sketched in Fig. \ref{fig:stereoscopic}:\commentout{Mazin:It might be good to introduce a theta2 plot? This would also help understanding that there is an irreducible background which has to be illuminated / subtracted on a statistical basis comparing ON and OFF regions. Carlos: shouldn´t this be part of the analysis chapter?  } since the development of the electromagnetic shower is symmetric around the incident direction of the primary gamma ray, the axis of symmetry of each registered image represents the direction projected into the FOV as observed from the position of each telescope. Actually, this projection, together with the position of the center of the telescope mirror, defines a plane in real space. If the telescopes are pointing to the same position in the sky, then the crossing point of these projections is, within the statistical and systematic errors of the determination of the images axes, the primary gamma-ray incident direction because it is the only direction common to all image axes. Additional information can be extracted if the description based in planes in real space is used, since the intersection of these planes does not only provides the direction, but also the impact point if the extrapolation of the trajectory to the ground. It is obvious that the accuracy of this determination then depends on the number of collected photons, which correlates with the energy of the primary gamma ray, and the number of telescopes that register the same shower. The resolution reached for a single gamma ray for the current generation of IACTs is the range 0.1-0.05$^\circ$ for a primary energy above few hundreds of GeV.

Figures \ref{fig:skymap} and \ref{fig:theta2} are typical results obtained with current-generation IACTs. The first one displays a distribution of reconstructed directions of $\gamma$-rays around a source and the latter one the distribution of angular distances between these reconstructed directions and the known or inferred position (so-called $\theta^2$) of the source. A cut of $\theta^2\lesssim$0.02 is typically applied to define the signal region and estimate the significance of the source detection. 

The reconstruction of the primary gamma-ray energy relies on two facts already described: i) the number of electrons and positrons of the shower is approximately proportional to the primary energy; ii) the Cherenkov photons reaching the ground are  nearly uniformly distributed within a radius of 130m of the impact point on ground. Based on these, the energy reconstruction relies on building a parameterised estimate of the primary energy given the impact parameter and the register number of photons of each image after cleaning, which are combined accounting for the statistical fluctuations. The parameterisation is built by means of complex Monte Carlo simulations of the shower development, based on models of the atmosphere and electromagnetic interactions(see \cite{corsika} for example), and the telescope response using ray-tracing codes and detailed descriptions of the photosensors and DAQ. 

As usual, there are more sophisticated methodologies which improve the accuracy and precision of the reconstruction of the primary properties at the cost of complexity and computational resources \cite{random forest,model analysis,3d reconstruction,dl}.

\subsection{Typical performance and scientific plots}

Key parameters characterizing the performance of IACTs are flux sensitivity, angular resolution, and energy resolution. Whilst angular resolution refers to the accuracy of measuring the incoming direction of the gamma ray, energy resolution refers to the accuracy of measuring its energy. Flux sensitivity is the minimum flux of a gamma-ray source that can be detected beyond a certain statistical significance, usually five sigma, in a given amount of time. 

\begin{figure}
    \centering
    \includegraphics[width=0.8\textwidth]{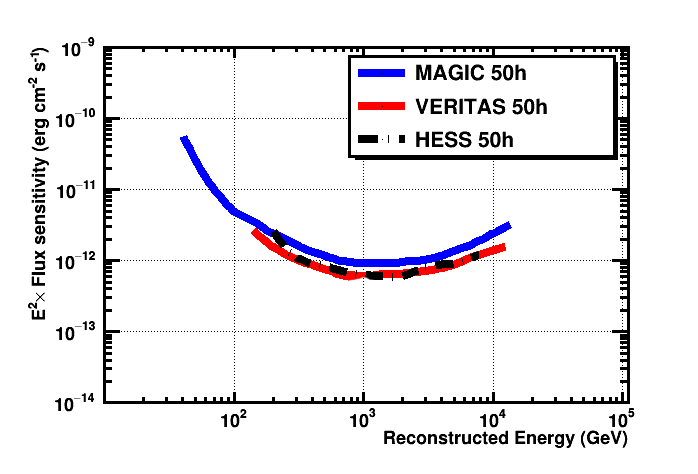}
    \caption{Sensitivity for several arrays of IACTs, that is, detectable flux with a statistical significance above the background fluctuations of five sigma in fifty hours. Extracted from \cite{MAGIC sensitivity}, \cite{VERITAS sensitivity} and \cite{HESS sensitivity}}
    \label{fig:sensitivity}
\end{figure}

\begin{figure}
    \centering
    \includegraphics[width=0.8\textwidth]{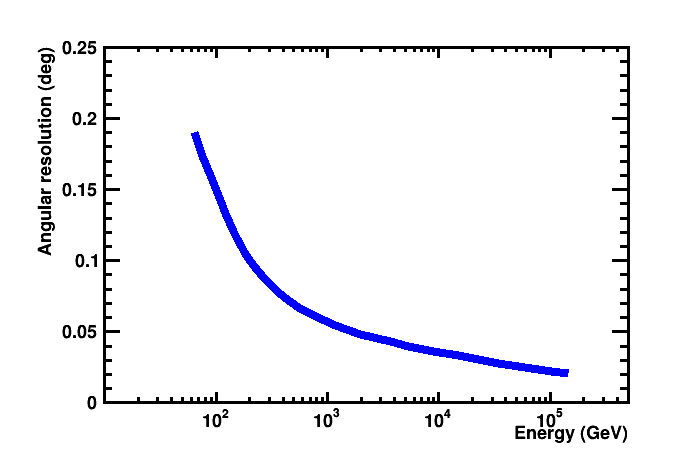}
    \caption{Angular resolution for a typical IACT array as a function of the gamma energy. Adapted from \cite{MAGIC sensitivity}.}
    \label{fig:a resolution}
\end{figure}

Performance figures are calculated by means of Monte Carlo simulations tailored to the observed properties of the detectors, resulting in typical angular resolutions in the range of 0.15 to 0.05 degrees and typical energy resolutions in the range of 10-15\%. This dependency on the energy of the gamma rays being observed is evident in figures \ref{fig:sensitivity} and \ref{fig:a resolution} for the sensitivity and angular resolution of some representative instruments.

\begin{figure}
    \centering
    \includegraphics[width=0.8\textwidth]{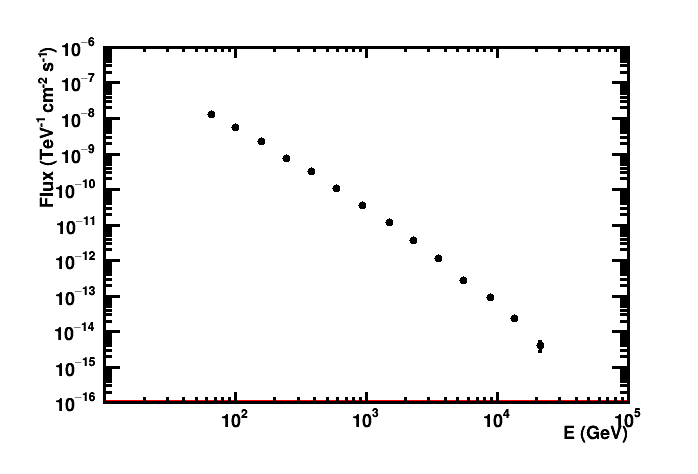}
    \caption{Reconstructed flux for the Crab Nebula with the MAGIC telescopes, adapted from \cite{MAGIC Crab}.}
    \label{fig:Crab sed}
\end{figure}

\begin{figure}
    \centering
    \includegraphics[width=0.8\textwidth]{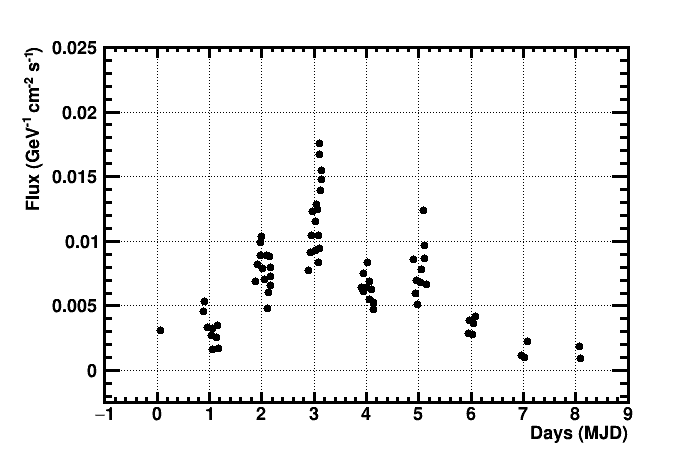}
    \caption{Typical light curve for a VHE bin, that is flux in the bin as a function of time, for a highly variable gamma-ray source. The x-axis corresponds to Modified Julian Date (MJD).}
    \label{fig:light_curve}
\end{figure}

Regarding typical results for IACTs, the reader can refer to \cite{handbook_analysis} for a comprehensive overview of high-level analysis techniques. Figures \ref{fig:skymap}, \ref{fig:Crab sed} and \ref{fig:light_curve} provide examples of what can be obtained through the analysis of a gamma-ray source with IACTs: they show respectively 1) the distribution of reconstructed directions of gammas in a given energy range; 2) the reconstructed energy spectrum for a bright source; and 3) the light curve, that is, the reconstructed flux as a function of time for a given energy range, for an exceptionally variable gamma-ray source.

\section{Current telescopes and future evolution of the technique}

Second generation IACT arrays such as H.E.S.S. in Namibia, MAGIC in Spain and VERITAS in USA have brought the technique to maturity and are still operational. The reader is referred to \cite{handbook_current_IACTs} for the status and a review of the roughly last 20 years of scientific results of these three instruments. 

The community behind these three IACT arrays have come together to design and build a full-sky open observatory called the Cherenkov Telescope Array Observatory (CTAO, \cite{handbook_CTA}). CTAO will consist of two arrays of IACTs in the northern and southern hemispheres.
CTAO-North will be located at the Roque de los Muchachos observatory (La Palma, Spain) and
CTAO-South at Cerro Paranal (Chile). The CTAO IACTs will have mirrors of three different sizes optimised for overlapping energy ranges: the Large-Sized Telescopes (LST) will be equipped with
the largest mirrors (23 m diameter) and target the lowest energies down to an energy of $\sim$20 GeV, the Medium-Sized Telescopes (MST) will be equipped with 12 m diameter mirrors and cover the range from roughly 100 GeV to a few TeV, and Small-Sized Telescopes (SSTs) with $\sim$4 m diameter mirrors will be sensitive to the highest energies up to hundreds of TeV.

CTAO is designed to operate for 30 years. Over this long period of time most probably the mechanics and optics of the CTAO telescopes will remain unchanged but the cameras will be upgraded with higher efficiency photodetectors, and faster trigger and readout electronics. Data analysis methods are expected to improve, especially with the application of new machine learning techniques, which may still be limited by computing power. 

With an increase in photodetection efficiency the IACT may expand to even lower energies, probably even below 10~GeV. Extracting physical information may prove challenging, given the fact that $\gamma$-ray showers develops farther and farther away from the telescope with decreasing energy and their images get correspondingly smaller and smaller. A low zenith 300 GeV $\gamma$-ray shower reaches its maximum at 8 km altitude above sea level while a 3~GeV shower does so at 14~km altitude. For higher zenith angles the distance to the shower maximum is even larger. Both camera pixelization and optical PSF will need to improve in order to resolve the relevant features of the shower image. 

The IACT technique may also evolve through innovative optics to increase the FOV, reduce the threshold or improve the optical quality \cite{cherenkov_schmidt, machete, plenoscope}. See \cite{handbook_future} for a more extensive review of future initiatives in ground-based gamma-ray astronomical detectors, including large FOV shower particle detector arrays.

%
%
%
%


\commentout{

\vspace{0.3cm}

{\bf Key-points to have in mind}:
\begin{enumerate}
\item The {\it Handbook of X-ray and Gamma-ray Astrophysics} is aiming to publish a work of tertiary literature, which provides easily accessible, digested and established knowledge derived from primary or secondary sources in the particular field. Therefore, your contribution should be clear and concise and be a comprehensive and up-to-date overview of your topic.
\item Length of text: every chapter normally consists of about 10,000-20,000 words (excluding figures and references), which would be about 20-40 typeset pages. However, this is not a rigid rule and it is something to discuss with the Section Editors of the Section of the chapter.
\item Colored figures are welcome in any standard format (jpg, tif, ppt, gif). If possible please provide the original figure in high resolution (300 dpi minimum). {\color{red}\bf Please do not forget to obtain permission in case the figure is from a published article}. For journals like ApJ, PRD, PRL, etc. you do not need the permission if you are an author of the article of the original figures, but for most journals you need the permission even if you are the author of those figures. However, if you create a new figures with minor changes, you can claim it is a new figure and you do not need any permission.
\end{enumerate}

\vspace{0.5cm}

\comment{Submission: 
\vspace{0.1cm} 

Please submit source files, compiled pdf file of the chapter, and figure permissions through the Meteor system.
\vspace{0.5cm}

arXiv policy:

\vspace{0.1cm}

Authors can post their chapter on arXiv if they wish to do it. In the comment field, please write something like ``Invited chapter for {\it Handbook of X-ray and Gamma-ray Astrophysics} (Eds. C. Bambi and A. Santangelo, Springer Singapore, expected in 2022)''. After the publication of the chapter, you may add the DOI on the arXiv page.

}
\newpage

\section{Tentative timeline}

This is our current timeline for the project:

\vspace{0.2cm}

\parbox[t]{3cm}{Mar-Sep 2021} 
\parbox[t]{15cm}{Authors write their chapters}

\vspace{0.2cm}

\parbox[t]{3cm}{Sep 2021} 
\parbox[t]{15cm}{Deadline for chapter submission}

\vspace{0.2cm}

\parbox[t]{3cm}{Sep 2021-Jan 2022} 
\parbox[t]{8.3cm}{Section Editors check the submitted contribution and recommend possible modifications\\ Authors amend and resubmit their chapter\\ Section Editors approve chapters for publications}

\vspace{0.2cm}

\parbox[t]{3cm}{Jan 2022} 
\parbox[t]{15cm}{We have all chapters ready for publication}

\vspace{0.2cm}

\parbox[t]{3cm}{Jun 2022}  
\parbox[t]{15cm}{Springer publishes the hard copy of the handbook}

\section{Cross-References \textit{(if applicable)}}
Include a list of related entries from the handbook here that may be of further interest to the readers.

\section{Note: References and Citations}
\subsection{References}
Should be restricted to the minimum number of essential references compatible with good scientific practice.\\
Include all works that are cited in the chapter and that have been published (including on the Internet) or accepted for publication. Personal communications and unpublished works should only be mentioned in the text. \textit{Do not use footnotes as a substitute for a reference list.}\
\subsection{Citations}
All references should be \textit{cited} in the text by the numbered style [n]; \cite{basic-contrib}, \cite{basic-online}, \cite{basic-DOI}, \cite{basic-journal} \cite{basic-mono}, etc. The recommended style for references is \textbf{Springer Basic Style} (examples are given below).

}
\end{document}